# Biology and Thermodynamics:
# Seemingly-Opposite Phenomena in Search of a Unified Paradigm

by

Shahar Dolev*

and

Avshalom C. Elitzur†

* The Kohn Institute for the History and Philosophy of Sciences, Tel-Aviv University, 69978 Tel-Aviv, Israel
  E-mail: shahard@ibm.net

† School of Physics and Astronomy, The Raymond and Beverly Sackler Faculty of Exact Sciences, Tel-Aviv University, 69978 Tel-Aviv, and The Seagram Center for Soil and Water Sciences, The Hebrew University, 76100 Rehovot, Israel.
  E-mail: cfeli@weizmann.weizmann.ac.il

It is probably not a coincidence that two of the pioneers of thermodynamics, Helmholtz and Mayer, were physicians. Thermodynamics studies the transformations of energy, and such transformations ceaselessly take place in all living systems (probably with important differences between the states of health and disease). Moreover, thermodynamics studies the elusive notions of order and disorder, which are also, respectively, the very hallmarks of life and death. These similarities suggest that thermodynamics might provide a unifying paradigm for many life sciences, explaining the multitude of life's manifestations on the basis of a few basic physical principles.

In this article we introduce some basic thermodynamic concepts and point out their usefulness for the biologist and the physician. We hope to show that thermodynamics enables looking at the riddles of life from a new perspective and asking some new fruitful questions.

## 1. The Second Law of Thermodynamics and its Bearing on Biology

Thermodynamics relies on three basic laws to study the transports of energy in physical systems and how they can be used to produce work. The First Law of Thermodynamics states that energy must be conserved. The Third Law states that it is impossible to reduce a system's temperature to the absolute zero. But the most interesting of the three is the Second Law. It states that within a closed system (that is, a system that no energy can enter or leave) entropy can only increase, or (when it is maximal) remain constant.

What is entropy? The dictionary tells us: "A measure of the unavailable energy in a closed system". There are several other, partly overlapping definitions of this important term. We will review them with the aid of the following simple example:

Imagine a sealed box divided in its middle by a partition. Let the right half of the box be in vacuum. If we puncture a hole in the partition, the gas will filtrate to the empty half until the entire box is equally full with gas. The filtration process increased the entropy of the system in the following senses:

1. **Equilibrium.** The initial state has low entropy since it was far from equilibrium (dense gas on one side, vacuum on the other). The final state is of high entropy since it has an even distribution of heat, pressure, etc.

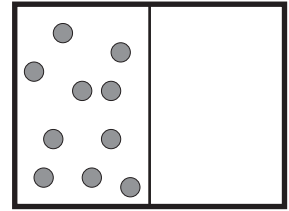

Figure 1 - Initial setup

2. **Bound energy.** Energy that can be used to do work is called "free energy" while energy that cannot be so used is "bound". In our example, free energy has degraded into bound energy. Suppose that the partition had been a piston. At the initial state, the pressure of the gas on the partition could do mechanical work. It was, therefore, free energy. At the final state, in contrast, all the energy has turned into chaotic, microscopic motions of the molecules that have spread all over the box. This energy can no longer be used for work [1] – another manifestation of entropy increase.

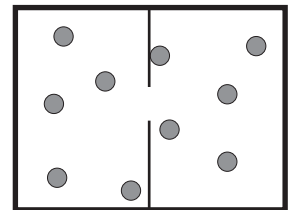

Figure 2 - After puncturing barrier

3. **Disorder.** Apparently, in our example, the final state, where the gas is equally dispersed in the box, is more ordered than the initial, unequal distribution of the gas. But actually it's the other way around. The "household definition" of order turns out to be consistent here with the physical one: house where the clothes, silverware, books, etc, are equally divided over the living room, kitchen, etc., is a house that leaves much to be desired. Order, therefore, is a state far from equilibrium.

4. **Irreversibility.** The spontaneous changes that the gas in the box underwent are irreversible. The likelihood that, by the same accidental motions of the molecules, all the gas will return by itself to the left half, is extremely low. Each gas molecule has a probability of 0.5 to be found in the right half. Since we are dealing with about $10^{25}$ particles (see section 3 below), the combined probability is $1/2^{10^{25}}$ (that's $10^{-10^{24}}$)! The degree of the unlikelihood for a system to return to its initial state is a measure of its irreversibility, hence of its entropy.

---

[1] An exception to this rule is the case where we use a colder environment outside the box. In this case the absence of equilibrium between the hot and cold reservoirs enables us to do work, just as at the box's initial stage. But then, of course, we are not dealing with a closed system, which is the case for which the Second Law holds.

5. **Number of microstates.** Another definition of entropy is based on the difference between the system's macroscopic and microscopic states. An ordered system allows only a small number of arrangements of its basic constituents. In contrast, there is a much larger number of arrangements that make an unordered system. In our case, there are much more possible arrangements of the gas molecules when the gas is evenly spread over the two halves, while the ordered state allows much fewer arrangements[2]. This insight is the basis of Boltzmann's definition of entropy, and here too, the household definition accords well with the physical one: There are only a few arrangements that make a house "ordered" and, unfortunately, numerous ways to make it disordered!

In summary, the Second Law states that entropy continuously increases. True, entropy can sometimes be decreased within a system, but only at the cost of energy investment that will increase entropy outside the system. And in this case, the system would not be a closed one. It would be *the system plus the environment* that constitutes a closed system, and in this closed system, again, the overall entropy has increased. To return to the household, you can make order in your house, but this will increase the entropy of your neighborhood. And if you make order in the neighborhood, you increase the entropy of your city. "You can't fight City Hall" is a common wisdom, and the Second Law seems to be the ultimate City Hall!

Having reviewed these definitions of entropy, it immediately strikes us that they also hint at some profound definition of the unique physical state we call "life", although in a very peculiar way. Notice, first, that the most illuminating demonstration of thermodynamics' pertinence to the life sciences comes from observing the processes to which the organism is subject upon *dying*. All the manifestations of decay that reduce the living tissue back into inorganic ashes share a fundamental physical characteristic, namely, complying with the Second Law: The decomposing organism goes back to the state of equilibrium (thermal, chemical, etc.) with its environment. Being alive, then, means being far from equilibrium with the environment, thereby manifesting the *autonomy* which is the very hallmark of life.

Another aspect that makes entropy the opposite of the living state has to do with the dynamic aspect of the Second Law, explained in the fifth definition above. Take, for example, a rolling ball on a rigid, flat surface. Initially, the ball harbors kinetic energy, but eventually friction will bring it to a halt. Where did the energy go? Since energy can never vanish (see the First Law) it can only change form. Tracing the "lost" energy, we will find that the ball has transferred its momentum to the molecules of the underlying surface and the ambient air. Doing so, it has lost kinetic energy while increasing the surface's molecules' thermal motion[3]. All in all, we can say that ordered energy – the

---

[2] See section 3.

[3] The rule is that temperature is actually a measure of the mean kinetic energy of the material's molecules. That is, the higher the temperature, the faster the molecules go (Sears, 1963).

*macroscopic* rigid-body motion of the ball – was transformed into disordered energy – the *microscopic,* thermal motions of multitude of surrounding molecules.

Here again, we can see the conversion of free energy to a bound one. The ball's original motion could have been harnessed to produce work (e.g. by turning a dynamo to generate electric current). However, the energy that was dispersed to the background environment cannot be used any more. The Second Law, that gives our world its time-arrow, is the reason why we *never* observe the opposite process: We won't believe a movie that shows a motionless ball beginning to roll spontaneously and then accelerating while the table cools down. We'll rather claim that the movie is running backward. But why is such a process impossible? After all, it does not violate the First Law, as the energy came from the microscopic motions of the surfaces and air molecules. Indeed, such a case is not absolutely impossible, but rather very, very unlikely: It would take more than the universe's lifetime for such an accident to occur somewhere. Practically, no one can trace these fractions of energy lost by the rolling ball and re-collect them back into a usable form. Even if such a method existed, it would end up consuming more energy than it has "freed."

In intriguing contrast, the living organism seems to exhibit exactly this impossible reversal. Magnasco (1993) has shown that under sufficient conditions, a biological microscopic "engine" is capable of drawing net motion from thermal energy alone. But we would like to point out that the living organisms can do much more. Take, for example, the muscles operation during bending of the arm: multitudes of microscopic muscle cells are cooperating by secreting, building and cross-linking actin and myosin filaments (Berne *et al*., 1993). Huge amounts of molecules move in a seemingly disordered manner, but somehow all these fractions of energy pile up to cause a *macroscopic,* ordered, motion of the arm. The percise microscopic control enables the muscles to reach maximum efficiency of 45% (Berne *et al*., 1993), as opposed to 25% efficiency in man-made engines (Sharpe, 1987). [4]

Even when no movement is apparent, the living body fights entropy all the time by performing enormous microscopic work: ion pumps keep the right concentration of ions across the cell membrane, various enzymes check cell structure and the DNA strands for errors, membrane proteins convey nutrients in and waste out, complex system cooperate to keep homeostasis, etc. It is these intracellular processes that later converge, with amazing precision, into macroscopic movements. We can therefore formulate a thermodynamic property that is unique to living systems:

> *In inanimate systems the microscopic motions are chaotic, resulting from the disintegration of the ordered motion of microscopic bodies. The living system, in*

---

[4] One might argue that the cooperation of many microscopic machines should be less efficient in comparison to one macroscopic machine, as the former case involves greater friction between the machines. Life, however, countered this problem by the highly ordered arrangement of the small machines so as to avoid friction. The muscle's molecules, for examle, are arranged along highly ordered polimers.

*contrast, maintains a very coordinated motion of its microscopic units, enabling them to converge at the right time into a macroscopic, ordered motion when needed.*

No less striking from the thermodynamic viewpoint is the course of development of a single creature, namely, its *ontogeny*. An oak tree, for example, begins its life as a zygote smaller than millimeter. Within a few years it consumes basic chemical elements from the surrounding air and soil, elements from highly *disordered* sources, only to organize them into the form of a mature, ordered, highly complex tree. Life has the unique ability to act against the normal course of events. Instead of scattering ordered motion of macroscopic objects into a multitude of tiny, disordered movements of microscopic molecules, living systems control the operation of single molecules, guiding minuscule amounts of energy and matter into an enormously ordered, macroscopic system.

Note that nothing of this violates the Second Law of thermodynamics. Living creatures are not closed systems, to which the Second Law applies. Since there is no free lunch in nature, living creatures must consume energy in order to create and maintain their internal order.

This is the answer given by all textbooks to the apparent contradiction between the Second Law and life's numerous manifestations. However, while this explanation is correct, it is highly insufficient. Nearly everything around us is an open system, and yet chairs and tables do not become alive. What is needed is a study of the particular processes by which very special and unique systems, namely, the living organisms, exploit their interactions with the environment in order to become more complex, ingenious and beautiful. In what follows we propose some guidelines for such a model.

## 2. Microstate vs. Macrostate

In the previous chapter we pointed out two scales by which one can look at a system. Let us examine these scales in more detail.

1. The microscopic scale, where one can examine the behavior of individual molecules.
2. The macroscopic scale, where one sees the overall state of the system, regardless of its individual molecules.

Thermodynamics taught us that it is not enough to look at the macroscopic level alone. One must take into account some properties of the microscopic level too. Consider, for example, the following experiment: There are two boxes, each with a string hanging out (Fig. 3). One box harbors a heavy rock connected to the string, while the other contains a spring connected to its string. When one pulls a box's string, he/she puts energy into the system. Although the two boxes look identical from the outside, there is a profound difference between their reactions to the pulling. Pulling the spring of the second box converts the energy into a usable, mechanical energy. This is a reversible process and the invested energy can be retrieved by letting the spring recoil. Pulling the rock within the other box will convert the energy to noise and heat, forms that are hardly usable. Only peering down to the molecular scale – i.e., studying the differences between the molecular structures of the rock and the spring – will reveal the difference between the two cases. When thermodynamics was constructed, it was realized that only one parameter is needed in order to describe the "usability" of the energy. That parameter is the entropy.

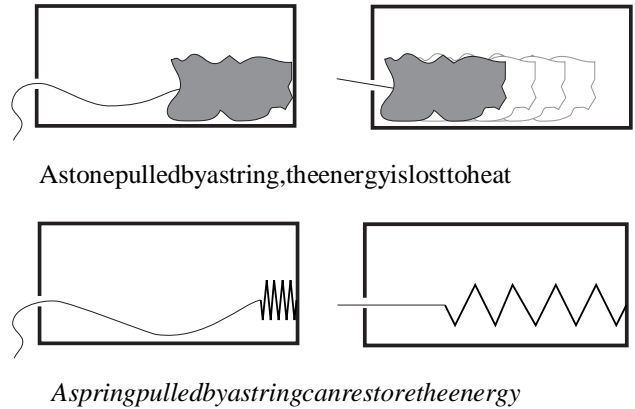

A stone pulled by a string, the energy is lost to heat

*A spring pulled by a string can restore the energy*

Figure 3 – Reversible vs. Irreversible processes

It was understood that one must consider the difference between what is visible to the naked eye on one hand, and the world of atoms and molecules on the other. The arrangement of a physical system at the macroscopic scale was named *macrostate*. A system's temperature or pressure are such macrostates. Now each such macrostate can be described by many different arrangements of the system's atoms and molecules. These arrangements in the microscopic scale were named *microstates*. In the previous section we have seen that high entropy is a macrostate that is compatible with many microstates, in contrast to the ordered state.

The biological significance of these formulations becomes conspicuous if we consider again the physical uniqueness of the living state. If we change the microstate of an inanimate object, say, a rock, by exchanging between the positions and momenta of some of its molecules, or even by replacing them with others, the rock will remain a rock; no difference will be noticed. Think, however, of an elephant or a whale: these are huge systems, but altering *their* microstates even slightly, by adding or subtracting a few grams of some hormone or neurotransmitter, could have drastic results – it may even kill the poor animal! Similarly, a single nucleotide in the DNA can have fatal consequences in most cases – or beneficial consequences in a few others. Such small may even change the fate of the entire biosphere. All living creatures, therefore, are unique in that they keep their inner autonomy by maintaining *Homeostasis*. In thermodynamic terms, organisms preserve their microstate. By using feedback loops, they keep their internal environment within those narrow required levels. We can therefore add another thermodynamic characteristic that is unique to the living state:

> *The living organism constantly resides in a macrostate that is compatible with a very narrow range of microstates, maintaining this improbable state as long as it is alive.*

## 3. The Phase Space

The thermodynamic explanation to entropy increase is a statistical one. To follow that explanation, we have to acquaint ourselves with the notion of *phase space*. This is a huge mathematical space, where each point can be assigned to a certain microstate of the entire system under examination. Actually the phase space has many dimensions [5], but as a model, a two-dimensional space is sufficient. The multi-dimensional structure of the phase space is such that when we map the different microstates of our system into it, all the states corresponding to a certain macrostate are adjacent. Thus one can divide the phase space into distinct regions corresponding to different macrostates.

Each point in the phase space describes *exactly* the positions and velocities of *all* the particles constituting our system. That means we can apply the laws of physics to predict how these properties would change once the system is at such a "point." The consequent microstates that would evolve from the initial one would be represented by new points in the phase space, arranged along a curve. Therefore, it is said that the system "wanders" through the phase space as time goes by.

As illustrated earlier by the household metaphor, there are very few ordered states, hence they occupy a very small region in phase space. The major part of this space (by several orders of magnitude) represents unordered states, *i.e.*, states of high entropy. This principle can be demonstrated by the partitioned box mentioned in Section 1. Following the puncture of the partition, each molecule of gas can be found anywhere within the container. That means that for each molecule, the volume that the system now takes in the phase state is twice as big (since the molecules can be found on a twice as large volume in the *x* direction). The phase space has a distinct set of dimensions for *each* molecule, hence the total volume that our system now takes is twice as big *for each additional particle*. Multiplying the contributions of all the gas molecules we get a factor of $2^n$, where *n* is the number of gas molecules.

For a 1-liter chamber, at 1 atmosphere and room temperature, we can calculate *n* using the classical equation for ideal gases:

$$P \cdot V = n \cdot R \cdot T$$

Where: *P=1 Atm., V=1 Liter, T=300 °K, R=1.362·10$^{-28}$ Liter·Atm/gm·deg*

We get: $n \cong 2.5 \cdot 10^{25}$

---

[5] There are six dimensions for *each* particle: three position dimensions and three of velocity.

That means that the volume our system now takes up in the phase space is approximately $2^{10^{25}}$ times bigger (that's more than $10^{10^{24}}$)! Since all microstates have equal probability to occur, the unordered state will have such a high probability that it is only natural to assume that the system will never return to the original, ordered, setup without an external aid. The classical thermodynamic argument states that if we leave the system alone for a long enough period, it might return one day to the original state. But you have to be *really patient* to see that, since the probability for such an event to occur is $1/10^{10^{24}}$!

According to the formalism of thermodynamics, entropy is proportional to the logarithm of the phase space volume, hence the entropy in the above case has increased by a factor of $10^{24}$. Now we can reformulate the Second Law in terms of the phase space: Even if a system begins at a very small region of the phase space that represent an ordered state, this region is surrounded by huge areas of unordered states. Left for itself, the system will most likely wander to these latter regions.

Applying this relation between micro- and macrostates to the life science, one can estimate the amount of order manifested by living systems. A protozoan (single-celled organism) would be highly unordered had its chemical composition been uniformly mixed. It is the unequal distribution of its enzymes, proteins, etc. between the protozoan's highly differentiated parts that makes it so ordered and capable of performing its unique biological tasks. A higher level of organization is manifested by the metazoan (multi-cellular organism), that have many types of differentiated cells and tissues, and yet a higher level is manifested by the ecosystem, where numerous different species maintain a highly complex web of dependencies.

So, looking around us, we can see that our planet has moved from an unordered state of an even mixture of chemicals, which prevailed four billion years ago, into the very ordered state that characterizes the biosphere today. Statistically, it seems, the odds for such a transition are nearly zero. Yet, the very fact that this statement is made by living creatures means that, long ago, the next-to-impossible *has* happened. Let us see how it actually took place.

4. **Life as an Information-Gaining Process**

We submit that life's secret in its battle against "all odds" lies in its ability to process information. The relation between entropy and information, long known to physicists, offers a very valuable insight for biologists. To grasp this profound relation, let us turn to the famous paradox associated with "Maxwell's Demon" (Leff & Rex, 1990).

We shall present the paradox by considering a setup similar to that considered in section 1 above, but Maxwell added a little twist to it. Suppose that, after the gas has spread to the entire box, we install a little door in the partition between the box's two halves, with a tiny demon guarding it (Figure 4). This demon is very smart. Whenever she sees a molecule of gas reaching from the right to the left half, she opens the door and lets the molecule pass through. But when a molecule tries to pass from left to right, she closes the door. The door is feather light and perfectly oiled, requiring very small amount of energy to open and close. As our demon continues with her work, she will eventually bring the system back to the original, low-entropy state (all the gas concentrated in the left half). This would be achieved with a negligible energy investment, hence with negligible entropy production outside the box. That is, the demon managed to decrease the entropy of our system by a factor of $10^{24}$ without paying the penalty to the universe's entropy. "City Hall" seems to have been defeated!

The paradox's solution is based on the concept of information: In order to let only the appropriate molecules pass and to stop the others, our demon needs information about them. It turns out that the amount of energy needed to identify the approaching molecule is such that it will soon create much more entropy than the order gained by this operation. [6]

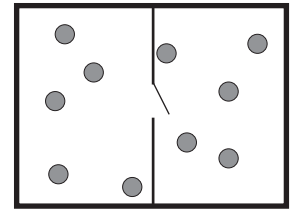
Figure 4 - Maxwell's demon

This paradox highlights the reciprocal relations between information and entropy, relations well known from computer science. Any generation, maintenance and processing of information take a proportionate cost in energy. Conversely – and this is a formulation of crucial importance – *the use of information allows saving energy*. If we have some information about a system, we can increase the system's order with only marginal waste of energy.

The relevance of this insight for biology is clear. The living cell must harness huge amounts of information for the purpose of fighting entropy. Using precisely crafted enzymes, the living cell is able to achieve high efficiency in its numerous biochemical operations. Each enzyme is a kind of a small Maxwell demon that uses the information gained during million of years of evolution to operate efficiently on it's substrate. This efficiency is beyond comparison to the efficiency that we humans achieve in designing machines and computers. Take, for example, sugar and other carbohydrates. Synthesizing them from their common constituents – water and carbon dioxide – lies, in principle, within the reach of modern technology. However, the cost of this production would be so high that no one would be able to buy these products. In annoying contrast, every grass leaf accomplishes this task every minute by using the negligible energy of little sunlight!

To take a more dramatic example, a tiger exerts enormous force to kill its prey. A Cobra, in contrast, kills its prey by merely spitting into its eye. What is appalling (or fascinating) in this act is the apparent disproportion between the force exerted on the prey and the fatal result. The choice of the appropriate neurotoxin, that matches the prey's synapses by its uncanny resemblance to its neurotransmitters, and the precise

---

[6] Maxwell's original example was slightly different in that there were equal amounts of gas in the box's two halves, with full equilibrium between them. The demon used the door to let only fast molecules to pass to one side and slow molecules to the other, until the gas was divided into a cold half and a hot half, in defiance of the Second Law. However, the essential physical points are the same in the original example and the one used above, as well as for the paradox's resolution.

"knowledge" of the location of a vulnerable point to penetrate the prey's vascular system – this is the information encoded in the Cobra's genes that allows it to save the energy that would be wasted by the tiger. But then, the gull's effortless gliding, the bee's honey production, the human's intelligence – in fact, every biological process – can be equally characterized by such Maxwell-demonic qualities.

Let us summarize. Adaptation, by definition, requires information about the environment to which the organism adapts. Natural selection is the process by which environmental information is incorporated into the species' genome. Once evolution is studied as a process by which organisms incorporate more and more information about their environment from generation to generation, the living organism appear as a very unique Maxwell demon that achieves incredible feats by a clever use of the thermodynamic affinity between information and energy. The magic formula is simple:

*Living organisms use little energy, but at the right place and at the right time!*

## 5. Complexity and the Struggle for Efficiency

So far, we have treated the living state as the mere opposite of the high entropy state. It would be mistaken, however, to simply equate "life" with "order." A third term, "complexity," is needed to capture the uniqueness of the living structure.

For an intuitive distinction between the three terms, think of three objects of the same size: a rock, a diamond, and a potato. The rock's entropy is the highest of the three – it is only an accidental assembly of minerals. The diamond, in contrast, is the most ordered object, as it is a perfect crystal of pure carbon. What about the potato? True, it is much less ordered than the diamond, yet it is far more *complex*. While it lacks the diamond's exact molecular structure and chemical purity, it is by no means as randomly assembled as the rock. The potato possesses, instead, highly detailed relations and correlations between its numerous constituents. Its cells resemble or complement one another to form well-defined tissues, and their dynamic operation reveals even more striking correlations. When we look at higher organisms, even at the simple level of their external form, this complexity becomes even more striking. Plants and animals are never perfect spheres, cubes or pyramids, yet they manifest clear symmetries and exact proportions between their different parts. We can say that complexity is a form of order, but of a very special kind: It is a structure whose parts are different from one another, yet they maintain very strict relations, both structural and dynamic, between them.

More precise mathematical formulations of complexity are discussed in detail elsewhere (Elitzur, 1998), but for our purpose the following observation suffices: Complexity, like order, cannot evolve spontaneously. On the contrary, it tends to degenerate into entropy just as order does. Similarly, its generation costs energy as the generation of order does. The living organism is clearly a very complex system. We therefore face an old problem in a new formulation: It is extremely unlikely that life on Earth evolved against the laws of thermodynamics, hence there must be some guiding principle that helped the biosphere to advance, against all odds, from the vast realms of disorder into smaller and smaller regions of growing complexity. That principle we are looking for must be powerful enough to create the magnificent, diverse, and perfectly adapted living creatures we see around us.

Our suggestion is that physicists are already familiar with that principle, yet have seldom noticed its relevance to biology. To comprehend this principle, let us think of evolution from the thermodynamic aspect of energy efficiency:

> *The ability of living systems to increase complexity is not accidental. Complexity is vital for efficiency. Life was therefore compelled to increase complexity as organisms fought for survival. The course of evolution can be rephrased as "Survival of the most efficient."*

The reason is simple: efficient organisms require less energy, thereby being able to survive tougher conditions (hunger, drought, etc.). As we saw in the section concerning information and efficiency, organisms had to accumulate information about their surroundings in order to achieve high efficiency. Only this way could they acquire the efficiency that enabled them to survive.

This trend can be demonstrated by the evolution of Hemoglobin (Lodish *et al*., 1995; Dickerson, 1983). Hemoglobin is highly adapted to its role, namely, transporting oxygen from the lungs to the cells. The hemoglobin molecule is a tetramer made of four sub-units, each capable of carrying one oxygen molecule. An energy barrier should be crossed in order to attach an oxygen molecule to each sub-unit. However, thanks to hemoglobin's unique structure, each oxygen molecule captured by it causes a geometric (allosteric) modification of the hemoglobin molecule, lowering the energy barrier.

The evolution of the hemoglobin molecule that has lead to its present efficiency can be traced by studying the molecule that performs the same task in more "primitive" species such as insects or cartilaginous fishes. It was found that the hemoglobin evolved out of a molecule that is similar to myoglobin (a molecule that transfers oxygen within the muscles). The myoglobin monomer is less efficient in carrying oxygen, having a higher energy barrier. Each sub-unit of the hemoglobin is a modified myoglobin molecule that was crafted during the evolution of vertebrates. In the course of evolution, in order to increase the efficiency of oxygen transfer, the simple myoglobin molecule was evolved to the more complex hemoglobin. The trend was driven by the need for higher efficiency, which was accomplished by incorporating information about the structure and physical qualities of the oxygen molecule. Complexity is the means by which efficiency was increased.

We began this section with an intuitive definition of complexity, but we should stress again that more objective measures have been proposed. Bennett (1988; Lloyd & Pagels, 1988) gave the following physical measure: Given the shortest algorithm for the construction of a certain structure, how much energy is needed for the computation of that algorithm so as to carry out the construction? Interestingly, both highly ordered and highly disordered structures turn out to have low complexity, while living organisms turn to have the highest complexity when taking into account the degree of computation needed to carry out the instruction of the organism's DNA.

Another approach has been adopted by Zotin and his co-workers (Zotin & Lamprecht, 1996, and references therein). Their work is based on the previously established relation between an organism's oxygen consumption and its bodily mass:

$$Q_{O_2} = aM^k,$$

where $Q_{O_2}$ is the oxygen consumption rate given in mW, $M$ is the organism's mass in grams and $a$ and $k$ are coefficients. They argue that there is a general trend in evolution that leads to increasing values of $a$. Indeed, comparative values of $a$ from a few main classes of animals accord with this claim. In other words, oxygen consumption per body mass increases with evolution, in accordance with the paleontological record. The data is admittedly very partial and insufficient, but the findings are exciting enough to warrant further study. They indicate that a simple thermodynamic measure might enable one to determine the degree of the organism's complexity.

## 6. The Molecular Scale

It seems that the high efficiency of living systems stems from their ability to control processes at the molecular scale, an accomplishment that no man-made machine has yet achieved. This unique ability of life to master microscopic mechanisms is, in fact, not so much of a surprise, since life *began* on the molecular scale. All life had later to do, then, was to keep its precious control at the molecular level. In other words, a disadvantage has been turned into an enormous advantage.

Let us describe this radical shift in more detail. By the simple laws of probability, life could not have begun at the macroscopic scale. The probability for even the tiniest bacteria to be spontaneously assembled out of an occasional binding of a myriad of wandering molecules is, of course, practically zero. However, the spontaneous assembly of a simple, self-replicating molecule is much more probable considering the time frame given for the emergence of life on Earth. The fact that life could only begin at the very simple microscopic level must have been a disadvantage for the first living systems, whatever they were. They were tiny, simple, and hence highly inefficient. However, this weakness eventually turned into an enormous advantage: control at the microscopic level was kept even when, by natural selection, macroscopic organisms evolved, granting living organisms the enormous efficiency that man-made machines are not even close to achieving today. As noted above, living organisms control chemical reactions at the single-molecule level, orchestrating the reactions of multitude of molecules to converge into macroscopic processes.

But why is efficiency greater when the system operates at the small scale? From the above thermodynamic formulations it follows that a process gets more efficient as it approaches *reversibility*. Perfectly reversible machines, though impossible in practice, are the most efficient ones. Now, again by the above formulation, machines approach reversibility the smaller they are.

Let us look at the molecular basis of this principle. Efficiency decreases when energy is lost to the environment in the form of random molecular motions (heat). What is unique about life is that the organism keeps energy loss low by controlling the processes at the molecular scale. When each molecule is directed to perform its specific task, only few can escape their destiny and lose energy to the surrounding environment. Compare this to man-made machines – let us take the extreme example of the most advanced, sub-micron computer chips: They rely on steering herds of electrons by macroscopic electromagnetic forces in the approximate direction. Inevitably, a great deal of them lose energy as they bump into one another, hitting other molecules in their vicinity and diverging from the intended direction. Only focusing the reactions to the single molecules or even single particles, as living organisms do, can minimize electron losses and increase efficiency.

It is even more instructive to compare the ordinary, wasteful technological process to one of the greatest wonders of animate nature, known as photosynthesis (Lodish *et al*., 1995). In this process photons are caught by the chlorophyll molecules, initiating a chain of reactions that transfers *single* electrons from one protein to another. At the end of the process several molecules of ATP and a single molecule of sugar are constructed. When humans tried to get energy from light by means of photoelectric cells, they ended up with a process similar to the micro-chip described earlier: A multitude of electrons that were popped from a semi-conductor by incoming photons are directed by electromagnetic force to the approximate direction. Electron motion over the semi-conductor is terribly wasteful, yielding an efficiency of only several percents. In order to achieve efficiency that equals that of plants, a pure crystal should be used, the production of which would cost thousands of dollars (Cheremisionoff *et al.*, 1978).

## 7. Biotechnology and Nanotechnology: Seeking the Efficiency of Living Systems

Admiring the incredible efficiency of living organisms, scientists are trying to exploit the latter's knowledge, acquired through billions of years of evolution, for technological purposes.

Nanotechnology is a new branch of technology that tries to achieve the efficiency of living organisms by reducing the machinery's scale. Nanotechnology's short-term goal is the production of micron sized machinery. The envisioned machines would be built by assembling single atoms and molecules together to form the desired precise structure. They will be able to replace us in unpleasant chores such as cleaning our environment, cultivating the ground and even medical tasks such as checking out our bodies and helping the immune system fight microbes and cancer (Feynman, 1960). Such a structure is said to be constructed *from the bottom up*.

The longer-term aspiration of nanotechnology is a generic *assembler* machine that will be able to build from the bottom up *any* product. Such an assembler will re-arrange single atoms and molecules so as to build the desired product. One might instruct the assembler to construct tasty *fillet-mignons* after emptying the garbage can into it. As unrealistic a sit sounds, this dream is perhaps not much different from the common feat of the growing oak tree mentioned earlier. Just as a tiny seed is able to collect minerals from the environment and rearrange them into living tissues, nanotechnology aspires to assemble a variety of products requiring only chemical ingredients, a construction program, and energy (Drexler, 1992).

Nanotechnology visionaries keep stressing the importance of operating at the small scale for increasing efficiency, by the precise control on each molecule and atom in the process. They rely on the natural examples we see around us as a proof for the viability of their master plan. They also consider thermodynamics when calculating energy intake, efficiency, and energy dissipation. Yet they neglect another point that is obvious from the thermodynamic point of view, namely, the fundamental relation between efficiency and information.

The biological structures and processes we see around us were crafted during billions of years. Each biochemical process in a living cell was programmed after evolution's trying an enormous number of different, random pathways. The process has gradually equipped the organism with invaluable information. In order to roughly asses the magnitude and value of this information, imagine the cost of a project whose aim would be to build a single ameba in the laboratory, out of the basic chemical elements. Any estimate would give a cost far above any nation's capabilities. The ameba, however, does it with infinitesimal costs every time it multiplies, by utilizing the information already stored in its DNA. Therefore, anyone who wishes to create a generic assembler that will be capable of producing *anything* overlooks the amount of information needed for such a project.

The prospect is much better, however, for a technology that seeks to exploit the information already encoded in the genomes of existing organisms. The myriad of species sharing our Globe, of which only a tiny fraction is known to science, stores an immeasurable treasure of pharmacological, agricultural and technological knowledge, only waiting to be studied. A technology that would take advantage of this treasure is certainly feasible.

## 8. Conclusions

In this article we have briefly discussed some points where thermodynamics offers fresh insights for the life sciences. New questions, ones that we did not even think about earlier, emerge when we look at the miracle of life from the thermodynamic perspective. While we are not sure about the answers, the questions themselves are important. Our aim has been only to appetize the medical and life scientist to become more acquainted with the growing literature dealing with this interdisciplinary field (Elitzur, 1994-1998 and references therein). We believe that the introduction of basic notions like entropy, information and complexity can add both depth and rigor to sciences as diverse as biochemistry, genetics, embryology, morphology and ecology.

Unfortunately, it is the latter field in which thermodynamic thinking yields the most far-reaching conclusions – and the ones that are ones most often ignored. Human societies keep ignoring the basic thermodynamic fact that any increase in a human's living standards entails a proportionate increase in the environment's entropy. Every member of Western society pollutes the environment with garbage, poisonous gases and heat to an extent that poses a serious threat to the entire biosphere. And on the top of it, mankind is recklessly multiplying, nearing the incredible figure of 12 billion predicted to populate the globe by the middle of the 21st Century. This expansion threatens to make all the achievements of modern medicine utterly impotent. No reasonable scenario allows such an explosion to happen without all the dire ecological consequences seen at the present – global warming, famines, diseases, etc. – becoming much worse.

Such calamities are inevitable consequences of the Second Law, to which most policy makers are totally oblivious. Not only do we pollute the globe with our ever-increasing waste products, we also directly ruin the biosphere's incredible complexity. Our generation witnesses one of the greatest extinctions of species that ever occurred on this globe. Biodiversity is rapidly shrinking in favor of the monotonous artificial environment that *Homo sapiens* creates everywhere, namely, the arrogant, human-centered blend of sky-scrapers, highways, malls, market-chains and their like. Konrad Lorenz (1974), the founder of ethology, a theoretical biologist and a physician by training, has once observed that the rapid expansion of human cities over the globe strikingly resembles the growth of a cancerous tumor. Indeed, in both cases complexity is ruined by the malignant takeover of only few of the living system's components. While genetic therapy seeks to combat cancer (by learning how to operate at its own small scale), we might be overlooking all along the very same calamity that we bring on the ailing tissue of which we are all part.

Many philosophers have objected to the attempts to explain biological phenomena by physical principles. "Reductionism" has become synonymous with disrespect for the phenomenon of life. In this paper we have tried to show that the contrary is the case. Not only does thermodynamics give a new dimension to the life sciences; it also emphasizes what we have intuitively known all along: That life is a state that is very unique, ill understood – and precious.

**Acknowledgments**

It is a pleasure to thank the Colton scholarship for its generous support.